\documentclass[final ]{aipproc}
\layoutstyle{8x11double}
\SetInternalRegister\hbadness{8000}

\def\pa{\partial}

\def\be{\begin{equation}}
\def\ee{\end{equation}}
\def\bea{\begin{eqnarray}}
\def\eea{\end{eqnarray}}
\def\nonu{\nonumber \\{}}
\def\half{{1 \over 2}}
\def\cq{{\cal{Q}}}

\def\a{\alpha}
\def\b{\beta}

\def\e{\epsilon}
\def\f{\phi}

\def\h{\eta}

\def\l{\lambda}
\def\m{\mu}
\def\n{\nu}
\def\o{\omega}
\def\p{\pi}

\def\x{\xi}

\begin{document}

\title
{Small Black Holes and Superconformal Quantum Mechanics}



\author{Joris Raeymaekers}{
  address={Korea Institute for Advanced Study, Seoul, Korea},
email={joris@kias.re.kr}, } \copyrightyear  {2005}

\begin{abstract}
Recently, Gaiotto, Strominger and Yin have proposed a holographic
representation of the microstates of certain $N=2$
black holes as chiral primaries of a superconformal quantum mechanics living on
D0-branes in the attractor geometry.
We show that their proposal can be succesfully applied to  `small' black
holes which are dual to Dabholkar-Harvey states and have vanishing horizon area in the leading
supergravity approximation.
\end{abstract}
\date{\today}
\maketitle
\section{Introduction}

The idea of holography, as developed by by 't Hooft and Susskind \cite{'tHooft:1993gx,Susskind:1994vu}, grew out of thinking about the physics of
four-dimensional black holes.
In string theory, the concept of holography has found a concrete realization in the $AdS/CFT$ correspondence. Therefore a natural setting
for the correspondence would be the charged extremal black holes
in four dimensions, whose near-horizon limit is $AdS_2 \times S^2$. Here, one would expect a holographic dual description of the near-horizon physics
in terms of  a conformally invariant quantum mechanics. However, this $AdS_2/CFT_1$ duality is much less well-understood than its
higher-dimensional counterparts.

Recently, a concrete proposal for such a holographic dual $CFT_1$ was made by Gaiotto, Strominger
and Yin (GSY)  \cite{Gaiotto:2004ij}.
Their proposal applies to a class of black holes in type IIA on a Calabi-Yau threefold
carrying D0 and
D4-brane charges.
The $CFT_1$ takes the form of a quantum mechanics of $N$ D0 brane probes moving
in the near-horizon geometry.
The super-isometry group of the background
 acts as a superconformal symmetry group on the quantum mechanics \cite{Claus:1998ts}.
The GSY proposal was that the black hole ground states should be identified with
 the chiral primaries of this quantum mechanical system.
Of particular importance were nonabelian $N$-D0 configurations corresponding to D2-branes
wrapping the black hole horizon and carrying $N$ units of worldvolume magnetic flux \cite{Simons:2004nm,Gaiotto:2004pc}.
These
experience a magnetic flux along the Calabi-Yau directions induced by the D4-branes in the background, and
chiral primary states correspond to lowest Landau levels. Their degeneracy was found to
 reproduce the leading order entropy formula but not the known subleading corrections to the entropy formula.
The GSY proposal realizes 't Hooft's old idea of having a finite number of degrees of freedom
per unit horizon area, although here it is the internal part of the horizon which gets partitioned into
lowest Landau level cells.

The analysis of GSY was performed for `large' black holes, which have a nonvanishing horizon
area in the leading supergravity approximation. Here the D4-brane charges $p^A$ are restricted
to obey the condition
$$ D \equiv {1 \over 6} C_{ABC} p^A p^B p^C \neq 0 $$ where $C_{ABC}$ are the triple intersection numbers on $CY_3$.
Furthermore, all $p^A$ have to be taken to be nonvanishing and large in order for $\a'$ corrections to the background
to be suppressed.

In the present note we find additional evidence for the GSY proposal by applying it to a different
class of black holes which have vanishing horizon area in the leading supergravity approximation,
but acquire a string scale horizon when higher derivative corrections are included
\cite{Sen:1995in,Dabholkar:2004yr}.
 For these `small' black holes,
the quantity $D$ vanishes.
We will limit ourselves to two special cases where the number of supersymmetries preserved by  the background
is enlarged and where the analysis becomes more tractable. We consider $D0-D4$ black holes in compactifications on
$T^2 \times M$, where $M$ can be either $K_3$ or $T^4$, and where the D4-branes are wrapped on $M$.
In both cases, there is a dual description of these black holes as BPS excitations in the fundamental
string spectrum, the so-called Dabholkar-Harvey states \cite{Dabholkar:1989jt}.
In contrast to the case of large black holes, only one of the magnetic charges $p^A$ is nonzero  and
 the D4-brane magnetic flux does not permeate all cycles
in the compactification manifold but  only has a component along $T^2$.

For more details and a more complete list of references we refer to \cite{Kim:2005yb}, of which this contribution is a summary.
\section{Quantum mechanics of the 2-charge black hole on $T^2 \times K_3$}
We consider type IIA compactified on $ T^2 \times K_3$ in the
presence of D0-branes and  D4-branes, the latter wrapped on the $K_3$. We
choose a basis $\{\o_A\}_{A = 1 \ldots 23}$ of 2-forms on $T^2
\times K_3$ in which $\o_1$ is the volume form on $T^2$ and
$\{\o_i\}_{i = 2 \ldots 23}$ are the 2-forms on $K_3$. The
4-dimensional effective theory is an $N=4$ supergravity theory. It
contains 24 homogeneous complex scalars $X^I,\ I= 0 \dots 23$ and
corresponding $U(1)$ gauge fields $F^I_{\m \n}$. Electric and
magnetic charges are labelled by integers $(q_I, p^I)$.

The D0-D4
system of interest carries nonzero $q_0$ and $p^1$, with all other
charges set to zero. In the supergravity approximation, the corresponding solution
has vanishing horizon area, but a horizon is generated when one includes  the leading 1-loop correction
to the prepotential.
The near-horizon geometry is  determined in
terms of the charges by the  generalized attractor
equations \cite{LopesCardoso:2000qm}. According to these equations, the number of D4-branes wrapping
a four-cycle determines the size of the dual two-cycle in the near-horizon region.
The resulting ten-dimensional type IIA background metric and RR fluxes
are given by
\bea
ds^2 &=& Q^2 \left( - r^2 {dt^2} + {dr^2 \over r^2}  + d \theta^2 + \sin^2 \theta d \f^2
\right)\nonu
&+& 2 dz d \bar z + 2 r g_{a \bar b} d z^a d \bar z^{\bar b}\nonu
F^{(4)} &=&  {p^1 \over 4 \p} \sin \theta d \theta d \f \wedge  \o_1; \qquad F^{(2)} =  {Q \over
g_s} dr \wedge dt\nonu
 \label{nearhor10d}
\eea
Here, we have chosen coordinates $(z, \bar z)$ on $T^2$ and
$(z^a, \bar z^{\bar a})_{a, \bar a=1,2} $ on $K_3$, and $g_{a \bar b}$ is proportional to the asymptotic Ricci-flat
metric on $K_3$.
In units in which $2 \p \sqrt{\a '} =1$, the radius
$Q$ of $AdS_2 \times S^2$ is given by
$Q = {g_s \over 2 \p} \sqrt{ p^1 \over |q_0|}$.
It's important to note that K\"ahler moduli of $K_3$ are not fixed to finite values at the horizon
but vary linearly with $r$. This is a consequence of the fact there are no D4-branes wrapped
on the cycles dual to the 2-cycles in $K_3$, hence the size of these cycles is not fixed by the attractor
mechanism. This constitutes the main difference with the large black hole case studied in \cite{Gaiotto:2004ij}, where all
internal four-cycles have a large number of D4-branes wrapped on them.

As in  \cite{Gaiotto:2004ij}, we will consider the quantum mechanics of a nonabelian
configuration of $N$  D0-brane probes
in the background (\ref{nearhor10d}) corresponding to a  D2-brane wrapping the horizon
$S^2$.
For large $N$, this system has an alternative description in terms
of a horizon-wrapping D2-brane with $N$ units of flux turned on on its worldvolume.
The target space seen by the brane is $R\times T^2 \times K_3$
with metric
$$ ds^2 = T \left( 2 Q d \x^2 + {\x^2 \over Q} dz d\bar z + {1 \over Q} g_{a \bar b} d z^a d \bar z^{\bar b}\right) $$
where we defined $\x \equiv 1/\sqrt{r}$ and $T$ is the mass of a horizon-wrapped D2-brane with N units of flux:
$ T = {2 \p \over g_s} \sqrt{(4 \p Q^2)^2 + N^2}.$
Note that the target space is in this case a direct product of $R\times T^2$ and $K_3$. This is a consequence of the
fact that the $K_3$ K\"ahler moduli vary linearly with $r$.
The bosonic Hamiltonian, to quadratic order in derivatives and in the limit $N \gg Q$, consists of
two decoupled parts:
\bea
H_{ bos} &=& H^{\ R \times T^2}_{ bos} + H^{K3}_{ bos}\nonu
H^{R \times T^2}_{ bos} &=& {1\over 8QT} P_\x^2 + {Q \over T \x^2} (P_z - A_z)(P_{\bar z} -
A_{\bar z}) + {32 \p^4 Q^5 \over g_s^2 T \x^2}\nonu
H^{K3}_{ bos} &=&{Q \over T} P_a g^{a \bar b} P_{\bar b}
 \label{bosham}
\eea
We have introduced a $U(1)$ gauge potential $A$ on $T^2$ obeying
$ d A = 2 \p p^1 \o_1$.

The full  quantum mechanics also contains sixteen fermions,
which are labeled as $(\l_\a, \bar \l_\a; \h_\a, \bar \h_\a; \h^a_\a, \h^{\bar a}_\a)_{\a = 1,2}$. These are roughly
the superpartners of
the bosonic coordinates $(\x;z ,\bar z;z^a, \bar z^{\bar a})$. The doublet index $\a$
indicates transformation properties under an $SU(2)$ R-symmetry which corresponds to spatial rotations.
The canonical anticommutation relations for the fermions are
\be
 \{ \l_\a,\bar \l_\b \} = \e_{\a \b};\ \ \
\{ \h_\a,\bar \h_\b \} = \e_{\a \b};\ \ \
 \{ \h^a_\a,\bar \h^{\bar b}_\b \} = \e_{\a \b} g^{a \bar b} \label{fermcomm}
 \ee

The (super-)isometries of the background ({\ref{nearhor10d}) act as symmetries on the quantum mechanics,
giving the symmetry group a superconformal structure \cite{Claus:1998ts}.
Due to the decoupling of the $R\times T^2$ and $K_3$
parts of the Hamiltonian, the symmetry group
naturally splits into a product $G_1 \times G_2$ with $G_1$ and $G_2$ acting on the $R\times T^2$ and $K_3$
parts of the wavefunction
respectively. It turns out that $G_1$ is the $N=4$ superconformal group $SU(1,1|2)_Z$ (where $Z$ indicates
the presence of a central charge) and $G_2$ is the supergroup of $N=4$ supersymmetric quantum mechanics (SQM).
The GSY proposal made in
\cite{Gaiotto:2004ij} states that the chiral primaries of the near-horizon D0-brane quantum mechanics
are to be identified with the black hole microstates. Applied to the case at hand, this means
that we have to count states of the form
$$|\psi \rangle \otimes |h\rangle$$
where  $|\psi \rangle$ is a chiral primary of
$SU(1,1|2)_Z$ and $|h\rangle$ is a supersymmetric
ground state of the $N=4$ SQM. We will now outline the construction of the states
$|\psi \rangle$  and $|h\rangle$.

The bosonic generators of of the group $SU(1,1|2)_Z$ acting on the $R\times T^2$
 Hilbert space consist of the Hamiltonian $H^{R \times T^2}$,
a dilatation generator $D$, a generator of special conformal transformations $K$ and
$SU(2)$ R-symmetry generators $T_{\a\b}$.
 The fermionic generators  $SU(1,1|2)_Z$  consist of supersymmetry generators $Q_\a, \bar
 Q_a$ and special supersymmetry generators $S_\a, \bar S_\a$. Their explicit expressions
 can be found in \cite{Kim:2005yb}.
Chiral primaries of $SU(1,1|2)_Z$ are characterized as follows \cite{Gaiotto:2004pc}.
We introduce the doublet notation
$$ Q^{++} = Q_1, \ \ \ Q^{-+} = Q_2, \ \ \  Q^{+-} = \bar Q_1, \ \ \ Q^{--} = \bar Q_2$$
and define
$$ G^{\a A}_{\pm \half} = {1 \over \sqrt{2}} ( Q^{\a A} \mp i S^{\a A} )$$
where $\a, A = +, -$. The relevant anticommutation relations are
\bea
\{ G^{\a A}_{\pm \half}, G^{\b B}_{\pm \half} \}&=& \e^{\a\b} \e^{AB} L_{\pm 1}\nonu
\{ G^{\a A}_{ \half}, G^{\b B}_{-\half} \}&=& \e^{\a\b} \e^{AB} L_0 + 2 \e^{AB} T^{\a\b} + \e^{\a\b} Z^{AB}\nonu
\eea
where $Z^{AB}$ is a c-number central charge matrix with $Z^{++} = Z^{--} = 0,\ Z^{+-} = Z^{-+} = 16 \p^2 Q^3/ g_s >0$.
The second anticommutator implies a unitarity bound
\be L_0 \geq j + 16 \p^2 Q^3/ g_s \label{bound}\ee
with $j$ the spin under the $SU(2)$ R-symmetry.
Primary states are annihilated by the positive moded operators  $G_{\half}^{\a A}$.
Chiral primaries $|\psi \rangle$ in addition saturate the bound (\ref{bound}),  hence they are also annihilated
by $G_{- \half}^{++}$:
\be G_{\half}^{\a A} | \psi \rangle = G_{- \half}^{++} | \psi \rangle =0. \label{cpcond}\ee

To construct the chiral primaries we use separation of variables into an $AdS_2$ and a $T^2$ component.
Denoting the $T^2$ component by $|\f\rangle$, it can be shown  that chiral primaries are in one-to-one
correspondence with states $|\f\rangle$ satisfying
\be
\h_\a (P_z - A_z)|\f\rangle = \bar \h_\a (P_{\bar z} - A_{\bar z})|\f\rangle = 0 \label{cond}
\ee
The equation $P_{\bar z} - A_{\bar z} = 0$ has no
normalizeable solutions, while the solutions to $P_z - A_z = 0$ are  the lowest Landau level wavefunctions $\f_k(z, \bar z)$.
The number of independent
lowest Landau level wavefunctions is given by an index theorem and is equal to the first Chern number
$${1 \over 2 \p}  \int_{T^2} d A = p^1.$$
Hence the equations (\ref{cond}) are solved by
\be
|\f_k\rangle = \f_k(z,\bar z) |0\rangle \label{T2sol}
\ee
where $|0\rangle$ is the vacuum state annihilated by the $\bar \h_\a$.
The resulting chiral primary states satisfying (\ref{cpcond})
are given by
\be
|\psi_k\rangle = \x^{- \half + {16 \p^2 Q^3 \over g_s}} e^{-2 QT \x^2} |\bar 0 \rangle \otimes |\f_k \rangle \label{primaries}
\ee
where $|\bar 0 \rangle$ is annihilated by $\l^{++}$ and $\l^{-+}$ and hence is bosonic under the rotational $SU(2)$. We have constructed
in this manner $p^1$ bosonic chiral primary states of \linebreak $SU(1,1|2)_Z$.

The $N=4$ SQM acting on the $K_3$ Hilbert space has  supersymmetry generators
$\cq_\a, \bar \cq_\a$
with anticommutation relations
$$ \{ \cq_\a, \bar \cq_\b \} = 2 \e_{\a\b} H^{K_3}; \qquad \{ \cq_\a, \cq_\b \} = 0 .$$

A supersymmetric ground state in the $N=4$ supersymmetric quantum mechanics  satisfies
$$ \cq_\a | h\rangle  = \bar \cq_\a |h\rangle =0 .$$
Such states are in one-to-one correspondence with the
Dolbeault cohomology classes on $K_3$. This can be seen
by using the well-known representation of  states $| h\rangle $ as differential forms on $K_3$.
In this representation, bosonic and fermionic states are represented by even and odd forms respectively.
The
supersymmetry generators are (up to  proportionality constants) identified with the Dolbeault operators
\bea
\cq_1 \leftrightarrow \pa &\qquad & \bar \cq_1 \leftrightarrow \bar \pa\nonu
\cq_2 \leftrightarrow \bar \pa^\dagger &\qquad& \bar \cq_2 \leftrightarrow  \pa^\dagger.
\eea
Since $K_3$ has 24 even harmonic forms, we find $24$ bosonic ground states $|h_M\rangle$ of the $N=4$ SQM.

To recapitulate, we have shown that the quantum mechanics of a single horizon-wrapping D2-brane
has $24 p^1$ bosonic chiral primaries $|\psi_k \rangle \otimes |h_M \rangle$ constructed
out of lowest Landau level wavefunctions on $T^2$ and harmonic forms on $K_3$.

We would now like to count the multi-D2-brane chiral primaries.
Note that the number of chiral primaries does
not depend on the background D0-brane charge $q_0$; hence for the purpose of counting ground states
we can take $q_0 \rightarrow 0$ and count the number of chiral primaries with total D0-brane charge $N$
in a background with fixed magnetic D4-charge $p^1$.
There is a large degeneracy of such states coming from the many ways the total number $N$ of D0-branes can be split into $k$ smaller
clusters of $n_i$ D0's such that
$$ \sum_{i = 1}^k n_i = N,$$
each cluster corresponding to a wrapped D2-brane that can reside in any of the $24 p^1$ bosonic chiral primary states.
The counting problem is the same as the counting the degeneracy $d_N$ of states at level $N$ in a $1+1$ dimensional CFT with $24 p^1$ bosons.
The generating function is then
\be
Z = \sum_N d_N q^N = \prod_n ( 1- q^n)^{-24 p^1}.\label{partK3}
\ee
This gives the asymptotic degeneracy at large $N$
$$
\ln d_N \approx 4 \p \sqrt{N p^1}
$$
which indeed equals to the known asymptotic degeneracy obtained from microscopic
counting \cite{Dabholkar:1989jt} or from the supergravity description incorporating
higher derivative corrections \cite{Dabholkar:2004yr}.

\section{Quantum mechanics of the 2-charge black hole on $T^6$}\label{T6}
We will now consider a compactification on $T^6$ with the same charge configuration as before,
i.e. a background produced by $q_0$ D0-branes and $p^1$ D4-branes wrapping a $T^4$. These are $1/4$
BPS black holes of 4-dimensional $N=8$ supergravity.  These black holes also have a vanishing horizon area
in the leading approximation and in this case also, it is expected that a horizon
is generated when including higher derivative corrections. Since all corrections to the prepotential vanish in this case,
the corrections that generate the horizon are expected to come from non-holomorphic corrections
to the supergravity equations, and it is not known how to incorporate these systematically at present.
We shall be a little cavalier  and simply  assume that the near-horizon limit of the quantum corrected background
is still of the form (\ref{nearhor10d}), with the $K_3$ metric now replaced by the  metric on $T^4$ and possibly with a different
value of the constant $Q$.

The resulting
quantum mechanics is then simply obtained by replacing the $K_3$ metric
$g_{a \bar b}$ in the previous section by the flat
metric on $T^4$. For the counting of chiral primaries, the only difference lies in the counting of ground states  of the
$N=4$ SQM, now corresponding to the Dolbeault cohomology of $T^4$. Since $T^4$ has 8 even and 8 odd harmonic forms,
we find   8 bosonic and 8 fermionic ground states. The counting problem for chiral primaries is now isomorphic to
counting the degeneracy at level $N$ of a CFT with $8 p^1$ bosons and $8p^1$ fermions. The partition function is
\be Z = \prod_n \left( {1 + q^n \over 1 - q^n} \right)^{8 p^1}. \label{partT6}\ee
This gives the asymptotic degeneracy
$$ \ln d_N \approx 2 \sqrt{2} \p \sqrt{N p^1} $$
which is in agreement with the known degeneracy from microscopic counting \cite{Dabholkar:1989jt}.

\section{Conclusion}
We have applied the GSY proposal to the case of two types of small black holes and found that the number of chiral
primaries agrees to leading order with the black hole entropy. For general $p^1$, the subleading terms
in the partition functions (\ref{partK3}), (\ref{partT6}) do not agree with the known corrections to the
entropy. It is amusing to note that, for $p^1=1$, the partition functions  (\ref{partK3}), (\ref{partT6}) do give the correct subleading terms.
The situation is similar to the problem of counting $D0-D4$ bound states at small string coupling, where the
correct counting involves taking into account nonabelian effects on the D4-branes. It will be interesting
to see whether the superconformal quantum mechanics picture can reproduce these subtleties.

\section*{Acknowledgements}

This work was done in collaboration with Seok Kim of KIAS, Seoul.
I would like to thank the organizers of PASCOS05, Gyeongju, Korea, where this
work was presented.

\end{document}